\documentclass[conference]{IEEEtran}
\IEEEoverridecommandlockouts
\usepackage{cite}
\usepackage{amsmath,amssymb,amsfonts}
\usepackage{graphicx}
\usepackage{booktabs}
\usepackage{siunitx}
\usepackage{xcolor}
\usepackage{tikz}
\usetikzlibrary{arrows.meta,patterns}
\usepackage[hidelinks]{hyperref}
\usepackage{colortbl}
\definecolor{headblue}{RGB}{221,235,247}

\IfFileExists{numbers.tex}{
\newcommand{\numEventsPerMin}{8.5}
\newcommand{\numBusShare}{41--42\%}
\newcommand{\numPedShare}{58--59\%}
\newcommand{\numSameCellLoss}{26--33}
\newcommand{\numOtherCellBus}{58--61\%}
\newcommand{\numOtherCellPed}{29--30\%}
\newcommand{\numOnsetBus}{0.47}
\newcommand{\numOnsetPed}{$\le$0.001}
\newcommand{\numBusPd}{0.94}
\newcommand{\numPedPd}{0.48}
\newcommand{\numFA}{3.8}
\newcommand{\numLeadBus}{83--90\%}
\newcommand{\numLeadPed}{53--65\%}
\newcommand{\numMapBus}{0.83 to 0.91}
\newcommand{\numMapPed}{0.28 to 0.45}

\newcommand{\numGhostGain}{0.83 to 0.88}
\newcommand{\numRefMargin}{35.9}
\newcommand{\numSNRreq}{9.3}
\newcommand{\numOverhead}{2.3\%}
\newcommand{\numAthreeTen}{3.58}

\newcommand{\numAthreeWideRef}{0.16}
\newcommand{\numGenieTen}{3.64}
\newcommand{\numOracleTen}{3.04}

\newcommand{\numAfiveTen}{3.51}
\newcommand{\numAfiveRef}{0.09}
\newcommand{\numCostOracleTen}{3.13}

\newcommand{\numSensePlanTen}{10.34}
\newcommand{\numSensePlanRef}{0.19}
\newcommand{\numTrueLossRef}{0.05}
\newcommand{\numTrueLossFrom}{15}
\newcommand{\numPairedMaxP}{0.084}
\newcommand{\numRelRedTen}{11\%}
\newcommand{\numTrueLossSigFrom}{20}
\newcommand{\numBEPosSigMax}{0.125}
\newcommand{\numBEUESigMax}{0.075}
\newcommand{\numTrueLossSigMaxP}{0.049}
\newcommand{\numCalPedAcq}{1.2}
\newcommand{\numCalPedOut}{0.9}
\newcommand{\numCalPosErr}{0.53--1.84}
\newcommand{\numGridCost}{1.9\%}
\newcommand{\numOnsetPedFast}{62\%}
\newcommand{\numErrMissTen}{+2.71}
\newcommand{\numErrFalseTen}{+0.70}
\newcommand{\numErrNoiseTen}{+1.30}
\newcommand{\numErrAllTen}{+4.92}
\newcommand{\numErrMissRef}{+0.108$^\dagger$}
\newcommand{\numErrFalseRef}{+0.006$^\dagger$}
\newcommand{\numErrNoiseRef}{+0.049}
\newcommand{\numErrAllRef}{+0.099}
\newcommand{\numDensLowTen}{1.76}
\newcommand{\numDensHighTen}{5.26}

\newcommand{\numAbsRedTen}{0.38}
\newcommand{\numRelRedRef}{62\%}
\newcommand{\numAbsRedRef}{0.057}
\newcommand{\numValueRange}{73--88\%}
\newcommand{\numValuePedRange}{48--65\%}
\newcommand{\numValueBusRange}{10--29\%}
\newcommand{\numValuePedTen}{53\%}
\newcommand{\numValueBusTen}{29\%}
\newcommand{\numPlanHorizon}{0.5}
\newcommand{\numSensePlanHO}{25.3}
\newcommand{\numSensePlanPP}{0.56}
\newcommand{\numAfiveHO}{6.6}
\newcommand{\numRunDuration}{60}
\newcommand{\numSpeedUE}{1.0 and 1.2}
\newcommand{\numDensCars}{4/12}
\newcommand{\numDensBuses}{1/2}
\newcommand{\numDensTrucks}{1/2}
\newcommand{\numDensSidePed}{6/12}
\newcommand{\numDensCrossPed}{2/4}
\newcommand{\numSpeedCar}{8--12}
\newcommand{\numSpeedHeavy}{6--10}
\newcommand{\numSpeedPed}{1.0--1.5}
\newcommand{\numOverheadDb}{0.26}
\newcommand{\numGeniePlanShareRange}{62--79\%}
\newcommand{\numGeniePlanShareTen}{73\%}
\newcommand{\numErrPerfTen}{+0.33}
\newcommand{\numErrPerfRef}{-0.042}
\newcommand{\numErrSizeTen}{+0.89}
\newcommand{\numErrSizeRef}{-0.018$^\dagger$}
\newcommand{\numErrAllUETen}{+6.91}
\newcommand{\numErrAllUERef}{+0.101}
\newcommand{\numErrRealTen}{+6.83}
\newcommand{\numErrRealRef}{+0.099}
\newcommand{\numMissPerFrameTen}{+0.87}
\newcommand{\numBEPosRange}{0.12--0.17}
\newcommand{\numBEUERange}{0.08--0.13}
\newcommand{\numBEPosRef}{0.64}
\newcommand{\numBEUERef}{0.19}
\newcommand{\numUAK}{8}
\newcommand{\numUAThetaTen}{0.3}
\newcommand{\numUAThetaRef}{0.1}
\newcommand{\numUAPlanIntervene}{6.6\%}
\newcommand{\numUAPlanTen}{+0.37}
\newcommand{\numUAPlanRef}{+0.008}
\newcommand{\numAfiveOHTen}{+0.37}
\newcommand{\numAfiveOHRef}{+0.002}
\newcommand{\numUAvsOHRef}{+0.006$^\dagger$}

}{}
\providecommand{\numEventsPerMin}{\prelim{8.5}}
\providecommand{\numBusShare}{\prelim{44--47\%}}
\providecommand{\numPedShare}{\prelim{53--56\%}}
\providecommand{\numSameCellLoss}{\prelim{26--34}}
\providecommand{\numOtherCellBus}{\prelim{68--71\%}}
\providecommand{\numOtherCellPed}{\prelim{27--34\%}}
\providecommand{\numOnsetBus}{\prelim{0.50}}
\providecommand{\numOnsetPed}{\prelim{0.01}}
\providecommand{\numBusPd}{\prelim{0.94}}
\providecommand{\numPedPd}{\prelim{0.46}}
\providecommand{\numFA}{\prelim{3.9}}
\providecommand{\numLeadBus}{\prelim{84--90\%}}
\providecommand{\numLeadPed}{\prelim{53--63\%}}
\providecommand{\numMapBus}{\prelim{0.82 to 0.90}}
\providecommand{\numMapPed}{\prelim{0.28 to 0.45}}

\providecommand{\numGhostGain}{\prelim{0.82 to 0.88}}
\providecommand{\numRefMargin}{\prelim{35.9}}
\providecommand{\numSNRreq}{\prelim{9.3}}
\providecommand{\numOverhead}{\prelim{2.3\%}}
\providecommand{\numAthreeTen}{\prelim{3.43}}

\providecommand{\numAthreeWideRef}{\prelim{0.14}}
\providecommand{\numGenieTen}{\prelim{3.48}}
\providecommand{\numOracleTen}{\prelim{2.87}}

\providecommand{\numAfiveTen}{\prelim{3.29}}
\providecommand{\numAfiveRef}{\prelim{0.08}}
\providecommand{\numCostOracleTen}{\prelim{2.97}}

\providecommand{\numSensePlanTen}{\prelim{9.07}}
\providecommand{\numSensePlanRef}{\prelim{0.13}}
\providecommand{\numTrueLossRef}{\prelim{0.04}}
\providecommand{\numTrueLossFrom}{\prelim{15}}
\providecommand{\numPairedMaxP}{\prelim{0.005}}
\providecommand{\numRelRedTen}{\prelim{10\%}}
\providecommand{\numTrueLossSigFrom}{\prelim{20}}
\providecommand{\numBEPosSigMax}{\prelim{0.125}}
\providecommand{\numBEUESigMax}{\prelim{0.075}}
\providecommand{\numTrueLossSigMaxP}{\prelim{0.049}}
\providecommand{\numCalPedAcq}{\prelim{1.2}}
\providecommand{\numCalPedOut}{\prelim{0.9}}
\providecommand{\numCalPosErr}{\prelim{0.5--1.8}}
\providecommand{\numGridCost}{\prelim{2.0\%}}
\providecommand{\numOnsetPedFast}{\prelim{62\%}}
\providecommand{\numErrMissTen}{\prelim{+0.68}}
\providecommand{\numErrFalseTen}{\prelim{+1.15}}
\providecommand{\numErrNoiseTen}{\prelim{+1.61}}
\providecommand{\numErrAllTen}{\prelim{+5.46}}
\providecommand{\numErrMissRef}{\prelim{+0.018}}
\providecommand{\numErrFalseRef}{\prelim{+0.053}}
\providecommand{\numErrNoiseRef}{\prelim{+0.046}}
\providecommand{\numErrAllRef}{\prelim{+0.071}}
\providecommand{\numDensLowTen}{\prelim{2.15}}
\providecommand{\numDensHighTen}{\prelim{4.44}}

\providecommand{\numAbsRedTen}{\prelim{0.32}}
\providecommand{\numRelRedRef}{\prelim{71\%}}
\providecommand{\numAbsRedRef}{\prelim{0.06}}
\providecommand{\numValueRange}{\prelim{70--89\%}}
\providecommand{\numValuePedRange}{\prelim{40--64\%}}
\providecommand{\numValueBusRange}{\prelim{11--32\%}}
\providecommand{\numValuePedTen}{\prelim{46\%}}
\providecommand{\numValueBusTen}{\prelim{32\%}}
\providecommand{\numPlanHorizon}{\prelim{0.5}}
\providecommand{\numSensePlanHO}{\prelim{23.7}}
\providecommand{\numSensePlanPP}{\prelim{0.54}}
\providecommand{\numAfiveHO}{\prelim{5.9}}

\newcommand{\figorboxw}[4]{%
  \IfFileExists{figs/#1}{\includegraphics[width=#2]{figs/#1}}%
  {\fbox{\parbox[c][#3][c]{#2}{\centering
   \textcolor{red}{Placeholder: \detokenize{figs/#1}. #4}}}}}
\newcommand{\figorbox}[3]{%
  \IfFileExists{figs/#1}{\includegraphics[width=\columnwidth]{figs/#1}}%
  {\fbox{\parbox[c][#2][c]{0.92\columnwidth}{\centering
   \textcolor{red}{Placeholder: \detokenize{figs/#1}. #3}}}}}

\begin{document}

\title{ISAC-Assisted Proactive Handover\\
in mmWave Networks: Value of Foresight\\
and Localization Requirements}

\author{
\IEEEauthorblockN{Seyed Bagher Hashemi Natanzi\IEEEauthorrefmark{1}\thanks{*Corresponding author: snatanzi@wpi.edu}, Bo Tang\IEEEauthorrefmark{2}}
\IEEEauthorblockA{\IEEEauthorrefmark{1}\IEEEauthorrefmark{2}Department of Electrical and Computer Engineering, Worcester Polytechnic Institute (WPI), USA}
\IEEEauthorblockA{Email: \{snatanzi, btang1\}@wpi.edu}
}

\maketitle
\maketitle
    \begin{tikzpicture}[remember picture, overlay]
        \node[anchor=north, yshift=-0.5cm] at (current page.north) {\fbox{\parbox{\textwidth}{\centering\small {\color{red}This work has been submitted to the IEEE Conference for possible publication. Copyright may be transferred without notice, after which this version may no longer be accessible.}}}};
    \end{tikzpicture}
\begin{abstract}
Millimeter-wave (mmWave) links can lose tens of decibels within milliseconds when a person or vehicle crosses the line of sight, and reactive handover recovers only after the outage has begun. Integrated sensing and communication (ISAC) promises
to let the network see blockers coming and move users to another cell
in advance. This paper asks how much such foresight is worth relative to
the best reactive handover, and what sensing accuracy is needed to
harvest it. We formulate cost-aware cell selection with knowledge of the
future signal-to-noise ratio as a sequential decision problem, define
the value of foresight through a policy-independent bound, and develop
an ISAC-assisted handover framework in which a radio unit senses
blockers with its own communication waveform and a simulated Open RAN xApp
plans handovers. Ray-traced experiments in a \SI{28}{GHz} street canyon,
reported on held-out scenarios, show that, with a short look-ahead,
perfect foresight closes most of the gap between service-aware (Layer-3)
reactive handover and an instantaneous oracle. Harvesting this value, however, requires the
user and nearby blockers to be localized to about a decimeter, well
beyond what our radar tracker and meter-level user positioning provide;
with realistic errors, even an uncertainty-aware planner largely defers
to reactive handover.
\end{abstract}


\section{Introduction}
The wide bandwidths of the millimeter-wave (mmWave) bands are central to
the capacity of 5G and 6G networks, but their short wavelength makes
links fragile: when a pedestrian or a vehicle crosses the line of sight
(LoS), the received power drops by tens of decibels within milliseconds
and stays low for up to seconds
\cite{rappaport2013mmwave,maccartney2017blockage}. Today's mobility
procedures are reactive. Beam failure recovery and handover are
triggered by degraded measurements, filtered over a time-to-trigger, so
the user experiences at least part of every outage. Outages dominate the
experience of latency- or rate-critical services.

Proactive mitigation requires knowing the blocker before it arrives.
Prior work has predicted blockage and triggered proactive handover from
cameras or dedicated radars
\cite{charan2021vision,demirhan2022radar,alkhateeb2023deepsense}, which
adds sensors, cost, and privacy concerns. Integrated sensing and
communication (ISAC) offers an alternative: the radio unit senses its
surroundings with its own OFDM waveform \cite{liu2022isac,sturm2011ofdm},
3GPP has studied ISAC use cases and standardized sensing-target channel
models \cite{3gpp22837,3gpp38901}, and Open RAN (O-RAN) lets an xApp
act on such information via the E2 interface
\cite{polese2023oran,bonati2021intelligence}.

Whether this pays off depends on two quantities that are rarely
separated. The first is the \emph{value} of foresight: what a network
with perfect knowledge of the future could gain over the best reactive
handover. The second is the \emph{realizable} part: what a practical
predictor and policy actually deliver. Existing studies typically
report gains of a specific predictor over link-failure recovery or
untuned triggers, so neither quantity is known, and it remains unclear
which sensing accuracy a network would need. Measuring the value
requires a strong reactive baseline and a policy-independent bound;
measuring the realizable part requires a model in which the blocker that
the radar observes is the same object that attenuates the link.

This paper addresses both questions. Our contributions are:
\begin{itemize}
\item A formulation of cost-aware cell selection with knowledge of the
future signal-to-noise ratio (SNR), solved exactly by dynamic
programming, which bounds any proactive policy and defines the value
of foresight relative to service-aware reactive handover (Layer-3 event
A5).
\item An ISAC-assisted handover framework in which a radio unit senses
blockers with its own waveform, tracks them with digital-twin-aware
processing, predicts each cell's SNR, and plans handovers in an O-RAN
xApp (modeled in simulation), including an uncertainty-aware planner.
\item A spatially consistent simulation built on Sionna~RT~2.2
\cite{hoydis2023sionnart,sionna} in which every moving body is both a 3GPP
radar target and a blocker of communication paths, released as open
source.\footnote{\url{https://github.com/CLIS-WPI/sense2connect}}
\item Held-out findings: which blockers matter, that foresight closes
most of the improvable gap with a short horizon and cost-aware
switching, and, by calibrated error analysis, the user and
blocker localization accuracy needed.
\end{itemize}

\section{Related Work}
\textit{mmWave blockage.} Measurements report deep and fast fading
caused by pedestrians at mmWave frequencies \cite{maccartney2017blockage},
and 3GPP models blockage with a stochastic model~A and a geometric
knife-edge model~B \cite{3gpp38901}. System-level studies mostly treat
blockers statistically, which decouples the blocker from anything the
network could observe.

\textit{Sensing-aided mobility.} Vision-aided methods predict blockage
and trigger proactive handover from camera images
\cite{charan2021vision}, and radar-aided methods predict beams and
blockages from a co-located automotive radar
\cite{demirhan2022radar,demirhan2022radarblockage}, supported by
multi-modal datasets \cite{alkhateeb2023deepsense}. RaDaR
\cite{alquraan2023radar} equips each small cell with a separate FMCW
radar, uses object, base-station and user heights to decide whether an
object obstructs the LoS, and predicts the time to blockage, with
gains over handover after link failure. Look Before
Switch \cite{li2025lookbefore} uses gNB-side ISAC to
track the vehicular user itself and triggers handover from its estimated
position and maneuver, cutting the handover interruption by more than
half relative to conventional handover.
We differ in three respects: we sense third-party blockers with the
communication waveform rather than the user or an auxiliary radar; we
compare against a tuned service-aware reactive policy (A5) and a
cost-aware oracle rather than link-failure recovery or fixed A3
settings; and we separate the value of foresight from the part that
sensing realizes. 

\section{System Model and Problem Formulation}
\begin{figure*}[t]
\centering
\begin{minipage}[c]{0.49\textwidth}
\centering
\begin{tikzpicture}[x=0.40cm,y=0.40cm,font=\scriptsize]
  \fill[gray!15] (-9,0) rectangle (10,-0.4);
  \draw[thick] (-9,0) -- (10,0);
  \fill[pattern=north east lines,pattern color=gray] (-9.6,0) rectangle (-8.6,9);
  \fill[pattern=north east lines,pattern color=gray] (9.6,0) rectangle (10.6,9);
  \node[below] at (-2.5,-0.4) {near lane};
  \node[below] at (2.5,-0.4) {far lane};
  \draw[fill=blue!25] (-3.5,0) rectangle (-1.5,1.6);
  \draw[dashed,red!70!black,thick] (-3.75,0) rectangle (-1.25,3.5);
  \node[red!70!black,left] at (-3.75,3.1) {truck};
  \node at (-2.5,0.8) {car};
  \draw[fill=blue!25] (1.5,0) rectangle (3.5,1.6);
  \fill (-7,1.5) circle (3pt); \node[below right] at (-7,1.4) {UE};
  \fill[black] (9.2,8) circle (3pt); \node[above left] at (9.2,8) {O-RU (facade, \SI{8}{m})};
  \draw[thick] (6.5,0) -- (6.5,5); \fill (6.5,5) circle (3pt);
  \node[below right] at (6.6,4.9) {lamppost \SI{5}{m}};
  \draw[thick,-{Latex}] (9.2,8) -- (-7,1.5);
  \draw[thick,dashed,-{Latex}] (6.5,5) -- (-7,1.5);
  \fill (-2.5,3.31) circle (1.6pt); \fill (-2.5,2.67) circle (1.6pt);
  \draw[-{Latex[length=1.5mm]},thin] (-4.4,6.2) -- (-2.6,3.4);
  \node[above,align=center] at (-4.4,6.1) {LoS over near lane:\\ \SI{2.7}{m} (lamppost), \SI{3.3}{m} (facade)};
\end{tikzpicture}
\\[1mm]{\footnotesize (a) Street cross-section}
\end{minipage}\hfill
\begin{minipage}[c]{0.49\textwidth}
\centering
\figorboxw{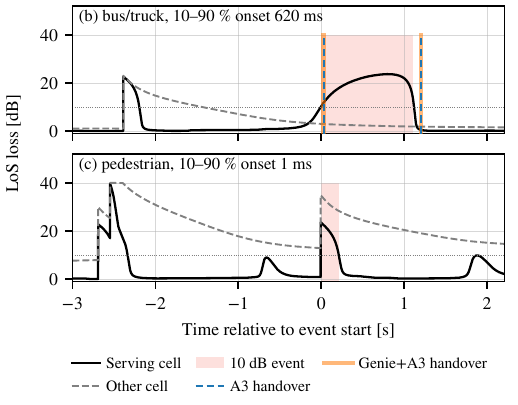}{3.45in}{6cm}{Onset panels.}
\end{minipage}
\caption{(a) Street cross-section: over the near lane the LoS passes above
cars but through tall vehicles. (b), (c) Model-B loss of an example
bus/truck and pedestrian outage (with its onset) at \SI{1}{ms}
resolution, with the handover instants of A3 and of the trigger genie
on top of A3 (Section~IV) at \SI{10}{dB} margin.}
\label{fig:scene}
\end{figure*}

\subsection{Network, Channel, and Blockage}
We consider a set $\mathcal{C}$ of cells, each served by an O-RAN radio
unit (O-RU) with an $N$-element array, and a single-antenna user
equipment (UE)~$u$ that can be served by any cell in $\mathcal{C}$. The
channel from O-RU~$c$ to the UE at frequency $f$ and time $t$ is
\begin{equation}
h_{c}(f,t)=\sum_{\ell} 10^{-L_\ell(t)/20}\,\alpha_\ell(t)\,
\mathbf{a}_c(\theta_\ell)\,e^{-j2\pi f\tau_\ell(t)},
\label{eq:channel}
\end{equation}
where $\alpha_\ell$, $\tau_\ell$, and $\theta_\ell$ are the gain, delay,
and departure direction of path~$\ell$ without blockers, $\mathbf{a}_c$
is the array response, and $L_\ell(t)=\sum_k L_{k,\ell}(t)$ is the loss that the moving bodies
$k$ cause on path~$\ell$, a function of their geometry relative to the
path segments. We use 3GPP blockage model~B \cite{3gpp38901}, in which
each body acts as a screen on every path segment it intersects, with
loss
\begin{equation}
L_k=-20\log_{10}\!\bigl(1-(F_{h_1}+F_{h_2})(F_{w_1}+F_{w_2})\bigr),
\label{eq:modelB}
\end{equation}
where each knife-edge term is
$F=\pi^{-1}\arctan\!\bigl(\pm\frac{\pi}{2}\sqrt{\pi\lambda^{-1}(D_1+D_2-r)}\bigr)$
and $D_1$, $D_2$, $r$ are the edge and direct distances of the segment;
losses add in decibels over bodies and segments. We denote by $\gamma_c(t)$ the SNR that cell~$c$ provides to the UE
under the receiver model at hand and by $\bar\gamma_c(t)$ its value
without blockers. The UE requires a rate
$R$, i.e., an SNR $\gamma_{\mathrm{req}}$, and is in \emph{outage}
whenever $\gamma_{c_t}(t)<\gamma_{\mathrm{req}}$ for its serving cell
$c_t$; every handover interrupts the user plane for $\tau_{\mathrm{HO}}$,
which also counts as outage.

\subsection{Sensing and Prediction Model}
The O-RUs transmit OFDM sensing symbols as part of their frames. With
$N_{\mathrm{sc}}$ subcarriers of spacing $\Delta f$ (bandwidth
$B=N_{\mathrm{sc}}\Delta f$) and a coherent processing interval of $M$
symbols spaced $T_s$, a monostatic radar resolves range, radial
velocity, and the maximum unambiguous velocity as
\begin{equation}
\Delta R=\frac{c_0}{2B},\qquad
\Delta v=\frac{\lambda}{2MT_s},\qquad
v_{\max}=\frac{\lambda}{4T_s},
\label{eq:res}
\end{equation}
where $c_0$ is the speed of light. Each moving body is a sensing target with state $\mathbf{x}_k(t)$
(position, velocity, class). The network obtains estimates
$\hat{\mathbf{x}}_k$ from its tracker and a UE position estimate
$\hat{\mathbf{p}}_u$ with errors, and predicts each cell's SNR in dB as
\begin{equation}
\hat\gamma_c(t)=\bar\gamma_c(t)-\hat L_c(t),
\label{eq:pred}
\end{equation}
where $\bar\gamma_c$ comes from the digital twin and the predicted
LoS-only loss $\hat L_c$ applies \eqref{eq:modelB} to the predicted
blocker trajectories on the line between O-RU~$c$ and $\hat{\mathbf{p}}_u$.

\subsection{Handover Problem and the Value of Foresight}
A handover policy selects the serving cell $c_t\in\mathcal{C}$ at every
step $t$ of duration $\Delta t$ and is judged by its outage time
\begin{equation}
J(\{c_t\})=\Delta t\sum_t \mathbb{1}\bigl\{\gamma_{c_t}(t)<\gamma_{\mathrm{req}}
\;\lor\; t\in\mathcal{I}(\{c_t\})\bigr\},
\label{eq:viterbi}
\end{equation}
where $\mathcal{I}$ contains the $\tau_{\mathrm{HO}}/\Delta t$ steps
after each switch, so that an interrupted step counts once.
\emph{Reactive} policies use only past measurements: A3 hands over when
the neighbor's filtered reference signal received power (RSRP) exceeds
the serving one by an offset plus hysteresis for a time-to-trigger
\cite{3gpp38331}; the Layer-3 event A5 hands over when the serving cell falls below one
threshold and the neighbor exceeds another, which makes it aware of the
service requirement.

The \emph{instantaneous
oracle} selects the better cell at every step without delay or
interruption. The \emph{cost-aware oracle} knows $\gamma_c(t)$ for all
cells and minimizes \eqref{eq:viterbi} exactly by the Viterbi algorithm
over the cell and the remaining interruption; no policy with any
predictor can do better. We call the outage difference between the best
reactive policy and the instantaneous oracle the \emph{improvable gap},
and the share of it closed by the cost-aware oracle the \emph{value of
foresight}. A \emph{grid-aligned} variant restricted to the decision
times of a practical controller isolates the cost of that restriction.

\section{ISAC-Assisted Handover Framework}
\textit{Sensing and tracking.} Per coherent processing interval, the
sensing O-RU forms range--Doppler--angle maps from the received echoes,
detects targets with two-dimensional cell-averaging constant false
alarm rate (CFAR) detection, rejects ghost
detections that mirror real targets on building walls known from the
digital twin (image method), clusters detections with DBSCAN, and tracks
them with an extended Kalman filter on range, angles, and radial
velocity. Optionally, tracks are constrained to lane center lines
(vehicles) and sidewalks (pedestrians, with a free model for crossings).

\textit{Prediction and planning.} An xApp in the near-real-time RAN
intelligent controller, modeled in simulation, receives tracks over E2
with period $T_r$ and acts after a loop delay $\tau_{\mathrm{E2}}$. At each report it predicts
$\hat\gamma_c$ by \eqref{eq:pred}; the \emph{planner} minimizes
\eqref{eq:viterbi} on $\hat\gamma_c$ over a horizon $H$ and its decision
grid and applies the first decision after the loop delay. The
\emph{genie-planner} uses the true future SNR and isolates the effect of
prediction errors. For comparison, \emph{trigger} policies hand over
when the serving cell's LoS is predicted blocked and the other cell's
clear.

\textit{Uncertainty-aware planning.} The \emph{uncertainty-aware planner}
keeps A5 and intervenes only if the expected outage
reduction, averaged over $K$ samples from the tracks' (inflated)
posterior and the UE position error, exceeds a threshold $\theta$.

\section{Experiments and Results}
\begin{figure}[t]
\centering
\figorbox{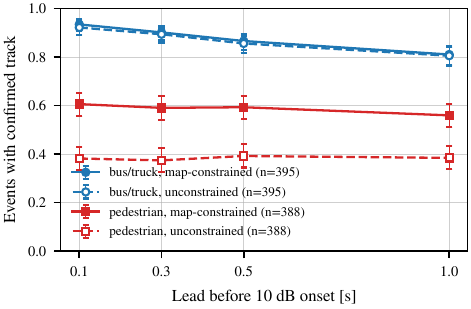}{3.2cm}{Share of 10 dB events with a confirmed track before onset.}
\caption{Share of outage events whose blocker is tracked a given time
before onset, map-constrained vs.\ unconstrained tracker (95\% Wilson
intervals; $n$: events in the cell of O-RU~0, pooled over mounts and densities).}
\label{fig:leadtime}
\end{figure}

\begin{figure*}[t]
\centering
\figorboxw{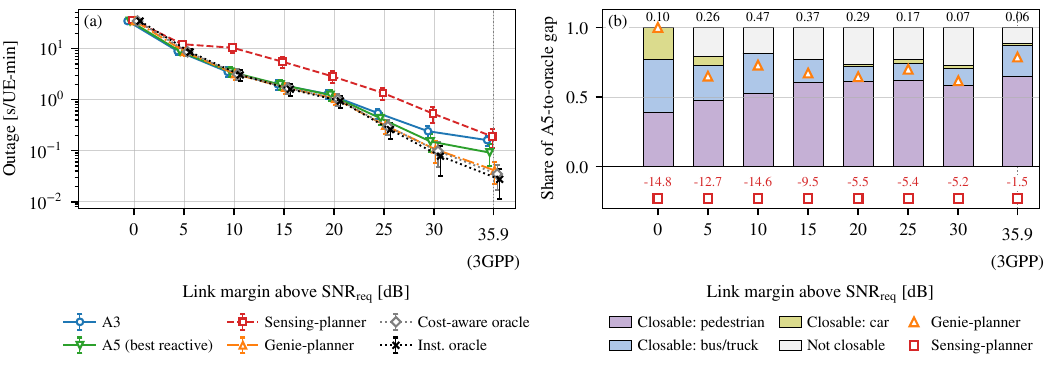}{\textwidth}{5cm}{Outage and value of foresight.}
\caption{(a) Outage time at \SI{400}{Mbit/s} versus link margin for A3,
A5, the sensing-driven planner, the genie-planner
($H=\numPlanHorizon$~s), and the two oracles (95\% CI over the 40 runs). (b) Share of the
A5-to-oracle gap closed by the cost-aware oracle, split by the class of
the blocker on the user's cell, with the genie-planner (triangles) and
the sensing-driven planner (below zero: worse than A5, in multiples of the gap); the
numbers above the bars: absolute gap [s/UE-min].}
\label{fig:results}
\end{figure*}

\subsection{Setup}
\textit{Scenario.} We simulate an \SI{80}{m} street canyon (periodic,
constant traffic density) at $f_c=\SI{28}{GHz}$ with two cells
(Fig.~\ref{fig:scene}(a)). O-RU~0 is on the north side, on a lamppost
(\SI{5}{m}) or a facade (\SI{8}{m}); O-RU~1 is a \SI{5}{m} lamppost on the
south sidewalk, half a block away. Each O-RU has an $8\times8$
half-wavelength array; UEs carry one antenna at \SI{1.5}{m} on the south
sidewalk. Traffic consists of cars ($5\times2\times\SI{1.6}{m}$), buses
($12\times2.5\times\SI{3.2}{m}$), trucks ($8\times2.5\times\SI{3.5}{m}$),
and pedestrians who walk the sidewalks or cross the street; every moving
body is a TR~38.901 sensing target \cite{3gpp38901} (five scattering
points for vehicles, one for pedestrians). Each \numRunDuration-s run
has two UEs (\numSpeedUE~m/s); low/high density places
\numDensCars{} cars, \numDensBuses{} buses, \numDensTrucks{} trucks, and
\numDensSidePed{} sidewalk and \numDensCrossPed{} crossing pedestrians
(\numSpeedCar, \numSpeedHeavy, and \numSpeedPed~m/s for cars,
buses/trucks, and pedestrians). Communication paths are
traced without bodies and attenuated by \eqref{eq:modelB} every
$\Delta t=\SI{10}{ms}$; only the onset analysis uses \SI{1}{ms}.

\textit{Sensing.} O-RU~0 operates as a monostatic radar under an ideal
full-duplex assumption, transmitting on one element and receiving on its
array; Sionna's RCS solver traces paths through the scattering points.
One sensing symbol per slot (numerology~3, 1024 subcarriers) and a
32-ms processing interval give $\Delta R\approx\SI{1.2}{m}$,
$\Delta v\approx\SI{0.17}{m/s}$, and $v_{\max}\approx\SI{21}{m/s}$ in
\eqref{eq:res}; one interval every $T_r=\SI{0.1}{s}$ occupies
\numOverhead{} of the resources. Sensing-based policies (the sensing,
true-loss, and uncertainty-aware planners and all conditions of
Table~\ref{tab:errors}) pay it as a rate loss, about \numOverheadDb\,dB on
$\gamma_{\mathrm{req}}$; A3, A5, the oracles, and the genie-planner do
not. The E2 loop delay is
$\tau_{\mathrm{E2}}=\SI{20}{ms}$ (modeled, without an implemented E2
service model), and the UE position, an exogenous input from UE-based or
network positioning (the radar senses only third-party bodies), carries
a Gaussian error of \SI{1}{m} per horizontal axis.

\textit{Link budget.} Transmit power (\SI{33}{dBm}) and UE noise figure
(\SI{13}{dB}) follow TR~38.802 \cite{3gpp38802}; unspecified losses
\cite{3gpp38830} are set to \SI{0}{dB}; the rate is Shannon's capped at the highest NR spectral efficiency
\cite{3gpp38214}. Here, $\gamma_c$ adds the path powers
with the full array gain after their model-B loss, an optimistic
approximation of the received multipath power, since no single beam
collects all paths with full gain. A rate of
\SI{400}{Mbit/s} requires $\gamma_{\mathrm{req}}=\numSNRreq$\,dB, and
$\tau_{\mathrm{HO}}=\SI{20}{ms}$. We define the link margin as the median
unblocked best-cell SNR minus $\gamma_{\mathrm{req}}$. The 3GPP budget
gives \numRefMargin\,dB on these short links; since longer links,
unspecified losses, or higher rates lower it, a common extra path loss
sweeps the margin from 0 to \SI{30}{dB}.

\textit{Protocol.} All policies are tuned per margin on five tuning
seeds (A3 over offsets and hysteresis up to \SI{15}{dB}; the tuned A5
time-to-trigger, \SI{40}{ms}, is the smallest value of its grid); methods were
fixed using ten development seeds, which also calibrate the tracker
error model below. All performance results come from a single final
evaluation on ten held-out seeds, each simulated for both mounts and
both traffic densities. Since the four runs of a seed share the traffic
and random draws, the seed, with the mean of its four runs, is the
statistical unit: we compare schemes pairwise with the Wilcoxon
signed-rank test over the ten seeds and give 95\% bootstrap CIs over
seeds for paired differences (Fig.~\ref{fig:sweep}); Fig.~\ref{fig:results}(a)
shows $t$-intervals over the 40 runs, Fig.~\ref{fig:leadtime} Wilson
intervals over events.

\subsection{Which Blockers Matter}
Users see about \numEventsPerMin{} LoS outage events
(\SI{10}{dB} model-B loss) per minute; Table~\ref{tab:classes}
summarizes them by class. Cars never cause them: for a UE at
height $h_u$ and an O-RU at height $h_o$ and horizontal distance $d$, the
LoS height at distance $x$ from the UE is
$z(x)=h_u+(h_o-h_u)x/d$, i.e., \SI{2.7}{m}--\SI{3.3}{m} over the near
lane: above a car, but through a bus or truck (Fig.~\ref{fig:scene}(a)).
Pedestrians block near the UE end of the ray.

\begin{table}[t]
\centering
\caption{Blockers of \SI{10}{dB} LoS outages (ranges: min--max over mounts; onset: median 10--90\% rise time; tracked: final map-constrained tracker, O-RU~0 events; value: share of improvable gap at \SI{10}{dB} closable per class, rest is not, Fig.~\ref{fig:results}(b)).}
\label{tab:classes}
\setlength{\tabcolsep}{2.2pt}
\setlength{\aboverulesep}{0pt}\setlength{\belowrulesep}{0pt}
\setlength{\extrarowheight}{1.5pt}
\begin{tabular}{lccccc}
\toprule
\rowcolor{headblue}
Class & Share of & 10--90\% & Other cell & Tracked & Share of \\
\rowcolor{headblue}
      & outages  & onset [s] & clear & 0.5 s ahead & value \\
\midrule
Bus/truck  & \numBusShare & \numOnsetBus & \numOtherCellBus & \numLeadBus & \numValueBusTen\\
Pedestrian & \numPedShare & \numOnsetPed & \numOtherCellPed & \numLeadPed & \numValuePedTen\\
Car        & 0\%          & --           & --               & --          & $\approx$0\%\\
\bottomrule
\end{tabular}
\end{table}

A blocker shadows the whole angular sector of the serving O-RU,
including the ground reflection: the best surviving path lies a median \numSameCellLoss\,dB below the
unblocked LoS, so the same cell keeps serving the UE only with a large
link margin, and otherwise only a second cell can. That cell
is clear in most heavy-vehicle outages but in few pedestrian outages,
because a pedestrian close to the UE often shadows both cells. The
outages also differ sharply in dynamics (Fig.~\ref{fig:scene}(b), (c)):
the loss caused by a bus or truck rises over \numOnsetBus\,s, whereas a
pedestrian next to the UE cuts the link almost at once:
\numOnsetPedFast{} of pedestrian onsets are shorter than \SI{10}{ms}.

\subsection{Sensing Performance}
With the detector tuned to about \numFA{} false alarms per processing
interval, the track detection probability is \numBusPd{} for buses and
trucks and \numPedPd{} for pedestrians (held-out seeds; per target
sample), and a confirmed track exists
\SI{0.5}{s} before onset in \numLeadBus{} of heavy-vehicle and
\numLeadPed{} of pedestrian outages (Fig.~\ref{fig:leadtime}). Of three
uses of the digital twin, static-clutter subtraction is equivalent to
blind zero-Doppler removal, because static clutter has exactly zero
Doppler in simulation; image-method ghost rejection removes 14--21\% of
the unmatched clusters and, at a fixed false-alarm budget, raises
heavy-vehicle detection from \numGhostGain{}; and, without ghost
rejection, map-constrained tracking raises detection from \numMapBus{} for heavy vehicles and from
\numMapPed{} for pedestrians. Sensing thus sees heavy vehicles reliably
and pedestrians only about half of the time: on the development seeds,
a pedestrian track is confirmed after a median of \numCalPedAcq~s and
then lost for \numCalPedOut~s at a time, and along-lane position errors
range from \numCalPosErr~m.

\subsection{The Value of Foresight}
Fig.~\ref{fig:results}(a) shows outage versus margin. A5, the best reactive policy on the tuning seeds, is our reference; on
the held-out seeds it is on par with the wide-grid A3 at low margins,
and its advantage grows with the margin (\numAfiveRef{} vs.\
\numAthreeWideRef~s/UE-min at the reference), since A3 switches on
RSRP differences even if the serving cell meets the service rate.

The cost-aware oracle closes \numValueRange{} of the improvable gap at
every margin from 5\,dB up (Fig.~\ref{fig:results}(b)); at \SI{10}{dB} it
reaches \numCostOracleTen{} vs.\ \numAfiveTen{} for A5 and
\numOracleTen{} for the instantaneous oracle. It does so largely by
making fewer, better-timed switches, so it avoids interruption as well as
blockage. In absolute terms, perfect foresight lowers the outage of A5
by \numRelRedTen{} (\numAbsRedTen~s/UE-min) at \SI{10}{dB} and by
\numRelRedRef{} (\numAbsRedRef~s/UE-min) at the reference; most of the
remaining outage occurs when no cell is usable. High traffic density
raises the outage about threefold (A5: \numDensLowTen{} vs.\
\numDensHighTen~s/UE-min at \SI{10}{dB}). The genie-planner closes
\numGeniePlanShareRange{} of the gap (\numGeniePlanShareTen{} at
\SI{10}{dB}); against the grid-aligned oracle, which exceeds the bound by
at most \numGridCost, a horizon of \numPlanHorizon~s suffices.
Trigger-based policies do not: even with perfect prediction, the trigger
genie stays at the level of the A3 it builds on (\numGenieTen{} vs.\
\numAthreeTen{} at \SI{10}{dB}), because blockage triggers ignore
sub-threshold losses and the cost of the switches they add. Foresight
thus needs cost-aware planning, but little look-ahead.

Pedestrian blockage accounts for \numValuePedRange{} of the improvable
gap and heavy vehicles for \numValueBusRange{}; cars contribute only at
margins of 0--\SI{5}{dB}, where their smaller losses already cause
outage. The blockers that sensing observes well, heavy
vehicles, cause slow outages that A5 follows reactively, while most of
the value lies in abrupt pedestrian blockage near the UE, which the
radar tracks in only about half of the cases.

\subsection{What ISAC Delivers}
With tracker predictions, the planner falls behind A5 at every
margin (Fig.~\ref{fig:results}; \numSensePlanTen{} vs.\ \numAfiveTen{} at \SI{10}{dB};
\numSensePlanRef{} vs.\ \numAfiveRef{} at the reference), with
\numSensePlanHO{} handovers per UE-minute against \numAfiveHO{} for A5
and a ping-pong rate of \numSensePlanPP: missed and false blockage
predictions translate into wrong switches. Fed with the true LoS loss,
while still paying the sensing overhead, the planner has a
lower mean outage than A5 from \numTrueLossFrom\,dB up
(\numTrueLossRef{} vs.\ \numAfiveRef{} at the reference); the advantage
is significant at \numTrueLossSigFrom--\SI{30}{dB} and at the reference
($p\le\numTrueLossSigMaxP$ each, without correction for the number of
margins) but not at \numTrueLossFrom\,dB ($p=\numPairedMaxP$).

\begin{table}[t]
\centering
\caption{Error budget: paired outage difference to A5 [s/UE-min] with
perfect tracks and with calibrated tracker errors injected; each ``+'' row adds one error type to perfect tracks,
and all values are total paired differences to A5; the size rule gives
cars bus size, so they are predicted to block the LoS
($^\dagger$: not significant, seed-level Wilcoxon $p\ge0.05$).}
\label{tab:errors}
\setlength{\tabcolsep}{4pt}
\setlength{\aboverulesep}{0pt}\setlength{\belowrulesep}{0pt}
\setlength{\extrarowheight}{1.5pt}
\begin{tabular}{lcc}
\toprule
\rowcolor{headblue}
Condition & \SI{10}{dB} & Reference\\
\midrule
Perfect tracks                    & \numErrPerfTen  & \numErrPerfRef\\
+ whole-track misses              & \numErrMissTen  & \numErrMissRef\\
+ persistent false tracks         & \numErrFalseTen & \numErrFalseRef\\
+ correlated pos./vel.\ error     & \numErrNoiseTen & \numErrNoiseRef\\
+ class-agnostic size rule        & \numErrSizeTen  & \numErrSizeRef\\
All tracker errors                & \numErrAllTen   & \numErrAllRef\\
All + \SI{1}{m} UE position error & \numErrAllUETen & \numErrAllUERef\\
\midrule
Real sensing-driven planner        & \numErrRealTen  & \numErrRealRef\\
\bottomrule
\end{tabular}
\end{table}

\textit{Error budget.} To attribute the shortfall, we inject our tracker's
errors into perfect tracks (Table~\ref{tab:errors}) with a model calibrated on the
development seeds: whole-track misses with the measured acquisition
delays and outage durations, persistent false tracks with the measured
rate, lifetime, and share of fragments next to real targets, position
and velocity errors along the lane that are correlated over time, and
the class-agnostic size rule of our predictor, which gives every vehicle
track the size of a bus. Missed
tracks are the largest single cause at every margin from \SI{5}{dB} up,
and their persistence matters: dropping detections independently per
frame costs only \numMissPerFrameTen~s/UE-min at \SI{10}{dB}, because the
tracker bridges single misses. All errors plus the \SI{1}{m} UE
position error reproduce the real planner's shortfall
(\numErrAllUETen{} vs.\ \numErrRealTen~s/UE-min at \SI{10}{dB}).

\begin{figure}[t]
\centering
\figorbox{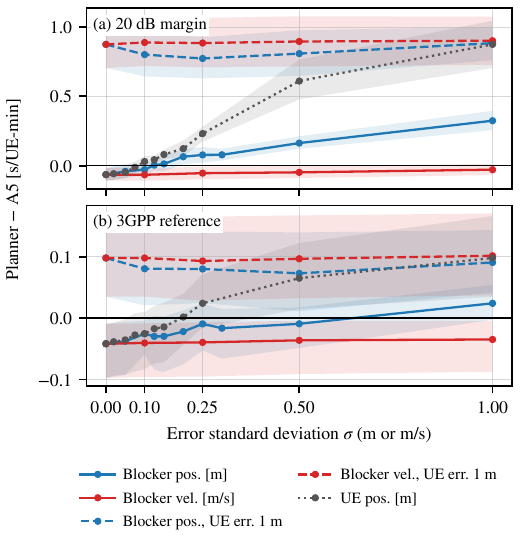}{3.8cm}{Planner minus A5 vs.\ error standard deviation.}
\caption{Accuracy requirements: outage of the planner minus that of A5
when one error source is added to perfect tracks, (a) at \SI{20}{dB} margin
and (b) at the reference (95\% CI). Below zero, foresight pays off.}
\label{fig:sweep}
\end{figure}

\textit{Accuracy requirements.} Fig.~\ref{fig:sweep} adds one error
source at a time to perfect tracks; $\sigma$ is the standard deviation
per horizontal axis (along the lane for blockers). Blocker velocity
errors up to \SI{1}{m/s} keep the mean outage below that of A5. For
margins of 15--\SI{30}{dB}, linear interpolation of the seed-mean
difference places the break-even points at a blocker position error of
\numBEPosRange~m and a UE position error of \numBEUERange~m
(\numBEPosRef{} and \numBEUERef~m at the reference). These are
single-error thresholds, not guarantees under simultaneous errors; as
exploratory evidence (pointwise tests without correction), the advantage
remains significant only up to \numBEPosSigMax~m and \numBEUESigMax~m,
respectively, at \SI{30}{dB} (less at lower margins). With a UE position error of \SI{1}{m}, no blocker accuracy
restores the advantage. A pedestrian next to the UE is about
\SI{0.5}{m} wide, so errors of this size move it in or out of the
predicted line of sight; in this deployment and error model, the user's
own position is the binding constraint.

\textit{Uncertainty-aware decisions.} Accounting for the uncertainty
does not recover the gain either. Tuned on the tuning seeds ($K=\numUAK$,
no covariance inflation, $\theta=\numUAThetaTen$~s at \SI{10}{dB} and
\numUAThetaRef~s at the reference), the uncertainty-aware planner never
intervenes at 10, 15, 20, and \SI{30}{dB} and in at most
\numUAPlanIntervene{} of the epochs otherwise. It is significantly behind
A5 (\numUAPlanTen{} and \numUAPlanRef~s/UE-min at \SI{10}{dB} and the
reference). At \SI{10}{dB} this deficit is entirely the sensing overhead
(\numAfiveOHTen), since it takes A5's decisions; at
the reference, the overhead accounts for \numAfiveOHRef{} and the
residual difference from overhead-matched A5 (\numUAvsOHRef) is not
significant. In our experiments, the tuned policy thus
rarely acts on the predictions.

\section{Discussion and Limitations}
In the studied deployment, most of the value of foresight lies in
pedestrian blockage close to the user. There, both the blocker and the
user must be localized to about a decimeter, more than a lamppost radar
and meter-level user positioning provide. Making ISAC pay off thus
requires precise user positioning and sensing geometries that see
the user's surroundings from close by, such as UE-side or bistatic
sensing.

The study has limitations: one street canyon with two cells; an O-RAN
loop modeled by its report period and delay rather than implemented,
e.g., on OpenAirInterface and FlexRIC \cite{nikaein2014oai,schmidt2021flexric}; monostatic sensing at one radio unit with
ideal full duplex and isotropic elements without a back plane, which
add mirror ghosts; model~B with opaque bodies, without full-wave
diffraction, window transmission, or user self-blockage; blockers at
constant velocity, which favors prediction, and ideal signaling without
link or handover-command failures, which favors reactive handover, and without
conditional handover or Rel-18 L1/L2-triggered mobility (LTM), which would strengthen reactive handover;
Shannon rates; a white Gaussian user position error; and an error model
calibrated to our tracker.

\section{Conclusion}
Perfect foresight can improve on service-aware reactive handover: a
planner that weighs the cost of switching closes most of the remaining
outage gap with less than a second of look-ahead. Turning this into
practice is difficult, because most of the gain comes from sudden
pedestrian blockage right next to the user. To capture it, the network
must locate the user and nearby blockers to about a decimeter. With
realistic radar tracking and meter-level user positioning, even an
uncertainty-aware planner falls back to reactive handover in our
deployment and error model. Knowing precisely where the user and its
surroundings are is therefore a key requirement for ISAC to help mmWave
mobility.

\section*{Acknowledgment}
This material is based upon work supported in part by NSF under Awards CNS-2120442 and IIS-2325863, and NTIA under Award No. 51-60-IF007. Any opinions, findings, and conclusions or recommendations expressed in this publication are those of the author(s) and do not necessarily reflect the views of the NSF and NTIA. AI assisted with coding and enhancing the text; the authors verified all content.

\bibliographystyle{IEEEtran}
\bibliography{refs}

\end{document}